\documentclass[twocolumn,english,prb,showpacs,superscriptaddress]{revtex4}
\usepackage{epsfig}
\usepackage{amsfonts}
\usepackage[figuresright]{rotating}
\usepackage{amssymb}
\usepackage{amsmath}
\usepackage{amsmath}
\usepackage{graphicx}

\begin{document}

\title{Topological phase transition in a generalized Kane-Mele-Hubbard model: A combined Quantum Monte Carlo and  Green's function study}
\author{Hsiang-Hsuan Hung}
\affiliation{Department of Physics, The University of Texas at
Austin, Austin, TX, 78712, USA }
\author{Lei Wang}
\affiliation{Theoretische Physik, ETH
Zurich, 8093 Zurich, Switzerland}
\author{Zheng-Cheng Gu}
\affiliation{Institute for Quantum Information, California Institute
of Technology, Pasadena, CA 91125, USA} \affiliation{Department of
Physics, California Institute of Technology, Pasadena, CA 91125,
USA}
\author{Gregory A. Fiete}
\affiliation{Department of Physics, The University of Texas at
Austin, Austin, TX, 78712, USA }
\begin{abstract}

We study a generalized Kane-Mele-Hubbard model with third-neighbor
hopping, an interacting two-dimensional model with a topological
phase transition as a function of third-neighbor hopping, by means
of the determinant projector Quantum Monte Carlo (QMC) method. This
technique is essentially numerically exact on models without a
fermion sign problem, such as the one we consider.  We determine the
interaction-dependence of the $Z_2$ topological insulator/trivial
insulator phase boundary by calculating the $Z_2$ invariants
directly from the single-particle Green's function.  The
interactions push the phase boundary to larger values of
third-neighbor hopping, thus stabilizing the topological phase. The
observation of boundary shifting entirely stems from quantum
fluctuations. We also identify qualitative features of the
single-particle Green's function which are computationally useful in
numerical searches for topological phase transitions without the
need to compute the full topological invariant.

\end{abstract}
\date{\today}
\maketitle

{\it Introduction.}-Recently, interest in a new state of matter,
topological insulators, has
exploded.\cite{hasan2010,moore2010,qi2011,fu2007prl,moore2007,roy2009}
$Z_2$ topological insulators (TI) do not require interactions for
their existence. However, intermediate strength electron-electron
interactions have been shown to drive novel phases in slave-particle
studies when the non-interacting limit is a
TI.\cite{Pesin:np10,Kargarian:prb11,Ruegg:prl12,Swingle:prb11,Maciejko:prl10,Kargarian13}
Interaction effects have also appeared in experimental studies on
the weakly correlated Bi-based TI.\cite{wangjian2011,liu2011}
Moreover, a recently discovered Kondo topological
insulator\cite{Kai2012} seems a promising venue to explore the
strongly interacting limit. An essential challenge in many-body
studies of TI systems is the direct characterization of the
interacting topological phases and phase transitions. This has
largely been accomplished with either mean-field-like approaches or
indirect evidence (such as the spontaneous appearance of an order
parameter, or the closing of the single-particle excitation gap).
Thus, it is important to understand the role of interactions in
topological phases beyond the standard independent-particle and
mean-field framework, ideally within an unbiased approach.

Various approaches, including the entanglement
entropy/spectrum\cite{turner2010,Kargarian:prb10} or K-matrix
theory\cite{lu2012} have also been proposed to characterize these
topological phases. In the case of $Z_2$ TI, topological invariants
can be identified in terms of the single-particle Green's
function.\cite{qi2008,Wang:prx12,Wang:prb12}  In certain cases, the
frequency domain-winding-number\cite{wanglei2011} and a
pole-expansion of the self-energy\cite{wanglei2012} could be useful
in identifying interaction effects in a topological phase
transition.  The pole-structure of the Green's function in dynamical
mean-field theory has been shown to be a  powerful tool in the study of
interaction effects in topological phases.\cite{ara2012} The
approach, however, still faces the limitation of being applicable only to local self-energy approximations.

Interaction induced topological phase transitions have been studied
in various models, including the Haldane-Hubbard
model,\cite{varney2010,varney2011} the Kane-Mele-Hubbard model
\cite{hohenadler2011,hohenadler2012,zheng2011,yu2011,Budich:prb12,Wu:prb12}
and the interacting Bernevig-Hughes-Zhang
model.\cite{wangleiepl2012,Tada:prb12,Budich12} Within these models,
there is also a topological phase transition at the single-particle
level even without interaction,\cite{Bernevig:sci06} which can be
induced by a staggered onsite energy,\cite{Kane:prl05} Rashba
spin-orbit coupling,\cite{Kane:prl05,kane2005a} or a third-neighbor
hopping, as we discuss in this Rapid Communication.  To study this
transition we use numerically exact determinant projector QMC to map
out the interaction dependence of the topological phase transition
as a function of third-neighbor hopping.  We find that interactions
tend to stabilize the topological phase, and we show the
zero-frequency behavior of the Green's function as a function of
third-neighbor hopping can be used to quantitatively determine the
phase boundary.

{\em Model.}-We consider the generalized Kane-Mele-Hubbard model
(KMH) on the honeycomb lattice (unit cell sites labeled A and B) with real-valued third-neighbor hopping $t_{3N}$: $H=H_0+H_U$
with
\begin{eqnarray}
  H_0 & = &
 - t\sum_{\langle i,j\rangle}\sum_{\sigma}c_{i\sigma}^{\dagger}c_{j\sigma}+i \lambda_{SO}\sum_{\langle\langle i,j\rangle\rangle} \sum_{\sigma}\sigma c^{\dagger}_{i\sigma}\nu_{ij} c_{j\sigma} \nonumber \\  & &  - t_{3N}\sum_{\langle\langle\langle i,j\rangle\rangle\rangle}\sum_{\sigma}c_{i\sigma}^{\dagger}c_{j\sigma} ,
\label{eqn:Ham}
\end{eqnarray}
and $H_U  = \frac{U}{2} \sum_{i} (n_{i}-1)^2$.  Here
$c^{\dag}_{i,\sigma}$ creates an electron with spin-$\sigma$ on
site-$i$; the fermion number operator is
$n_i=\sum_{\sigma}c^{\dag}_{i,\sigma}c_{i,\sigma}$; $\sigma$
runs over $\uparrow$ and $\downarrow$. The spin-orbit coupling strength is $\lambda_{SO}$, and
$\nu_{ij}=+1$ for counter-clockwise hopping with $\nu_{ij}=-1$
otherwise.\cite{Kane:prl05} The spin-orbit coupling term opens a
bulk gap and drives the system to a $Z_2$ TI for
$t_{3N}=0$.\cite{Kane:prl05}

\begin{figure}[t]
\epsfig{file=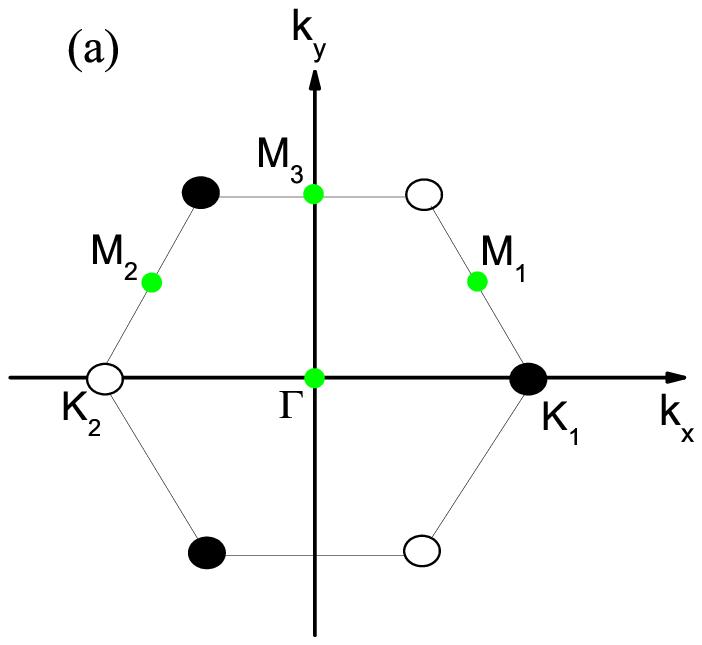,clip=0.1,width=0.53\linewidth,angle=0}
\epsfig{file=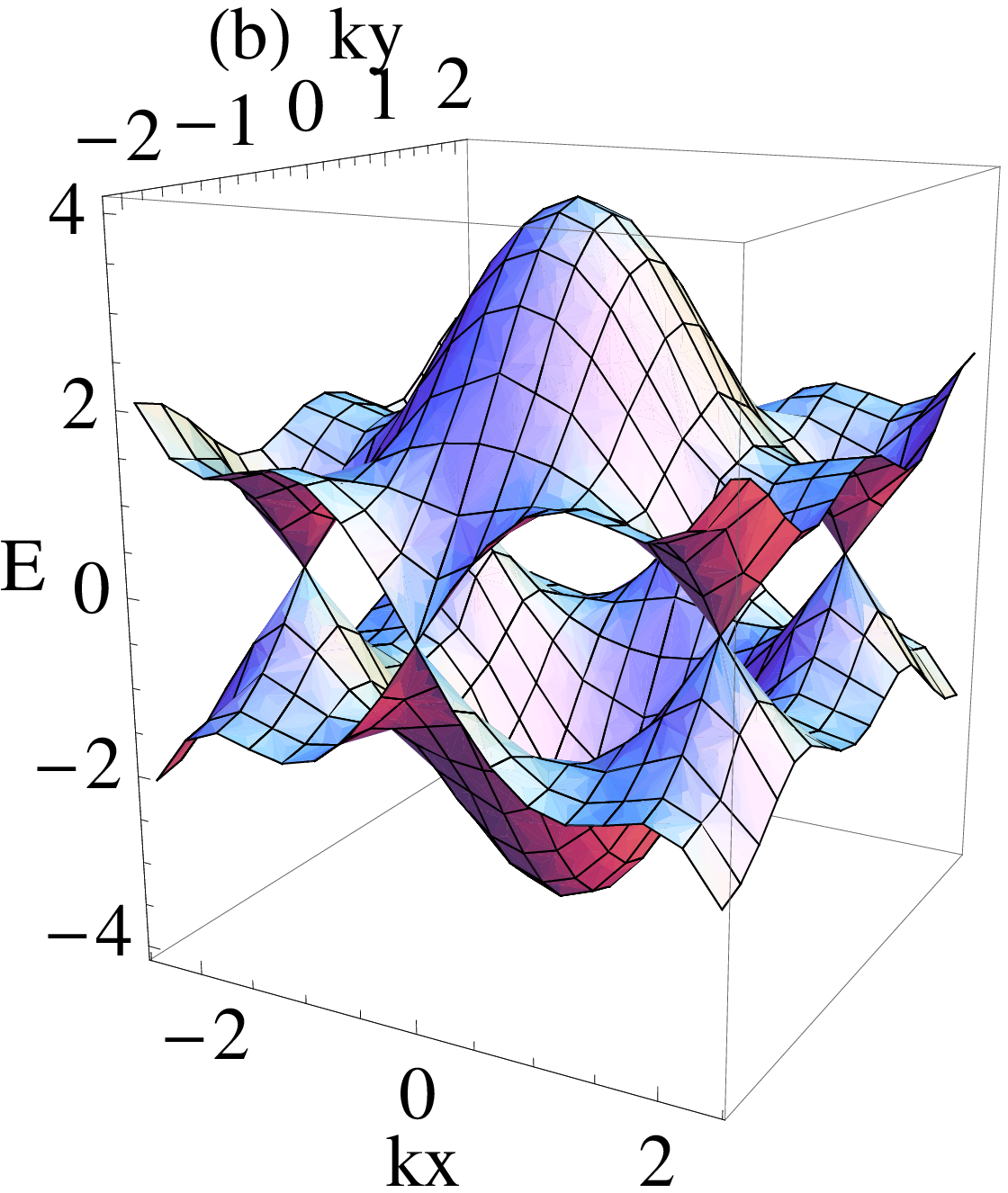,clip=0.1,width=0.43\linewidth,angle=0}
\caption{(Color online) (a) The first Brillouin zone of the
honeycomb lattice. For $t_{3N}=0$, the Dirac points are located at
$K_{1,2}=(\pm\frac{4\pi}{3\sqrt{3}a},0)$ labeled by open and solid
circles, respectively. The time-reversal invariant momentum (TRIM)
points labeled by the green dots are $\Gamma=(0,0)$,
$M_{1,2}=(\pm\frac{\pi}{\sqrt{3}a},\frac{\pi}{3a})$, and
$M_{3}=(0,\frac{2\pi}{3a})$.  (b) The noninteracting band structure
of the generalized KM model Eq.\eqref{eqn:Ham} at
$t_{3N}=\frac{1}{3}t$ (here using $\lambda_{SO}=0.3t$). Note that at
the critical point separating the trivial insulator from the TI the
Dirac cones shift to the TRIM points $M_{1,2,3}$, instead at
 $K_{1,2}$. } \label{fig:bandstructure}
\end{figure}

The Brillouin zone (BZ) of the honeycomb lattice  is shown in Fig.
\ref{fig:bandstructure} (a). For general $t_{3N}$ but vanishing
$\lambda_{SO}$, the model still exhibits a graphene-like band
structure with gapless Dirac cones located at
$K_{1,2}=(\pm\frac{4\pi}{3\sqrt{3}a},0)$, where $a$ is the lattice
constant. However, an arbitrary $\lambda_{SO}$ will open a bulk gap
and the generalized KM model turns into a $Z_2$ TI or a trivial
insulator depending on values of $t_{3N}$. We find that for $U=0$
the critical value of the third-neighbor coupling $t_c$ is
$t_{3N}=\frac{1}{3}t$. At $t_c$, the bulk gap closes and the gapless
Dirac cones shift away from the $K$ points and move to time-reversal
invariant momenta (TRIM), $M_{1,2}=(\pm \frac{\pi}{\sqrt{3}a},
\frac{\pi}{3a})$ and $M_3=(0,\frac{2\pi}{3a})$. The band structure
at the topological critical point is depicted in Fig.
\ref{fig:bandstructure} (b). As $t_{3N} <\frac{1}{3}t$, the system
is  a $Z_2$ TI, whereas as $t_{3N} > \frac{1}{3}t$ it is a trivial
insulator. At the noninteracting level, the value of $t_{c}$ is
independent of $\lambda_{SO}$.

We next consider the Hubbard interaction $H_U$, given below
Eq.(\ref{eqn:Ham}).  In the presence of the Hubbard interaction, the
topological phase boundary, $t_c$, shifts; a mean-field approach is
unable to accurately determine $t_{c}$ for $U\ne 0$. In fact, we
have verified that Hartree-Fock theory\cite{rachel2010} predicts no
shift at all for $U$ sufficiently small to avoid the magnetic
transition.  For $U$ larger than this critical value, $U_c$, the
topological band insulator state breaks down to a topologically
trivial magnetic
state.\cite{hohenadler2011,hohenadler2012,zheng2011,yu2011,Budich:prb12,Wu:prb12}
Since the generalized Kane-Mele Hubbard model we consider with the
$t_{3N}$ term still preserves the essential band features of the
Kane-Mele model, one can expect that in the strong coupling limit
$U>U_c$ our generalized model will also have a phase transition from
the $Z_2$ TI to the magnetic state.

To study physics not captured within a mean-field theory, we choose
a  moderate Hubbard interaction $U$ relative to the bandwidth (small
enough to avoid inducing the magnetic phase in the thermodynamic
limit). Our main goal here is to demonstrate how the single-particle
Green's functions computed within QMC in a fermion sign-free problem
can be used to identify a correlated TI phase and topologically
trivial insulating state.  We leave a detailed analysis of the large
$U$ case for a future publication. At half-filling, \textit{i.e.,\/}
one fermion per site, the system has a particle-hole symmetry and
the QMC simulations can perform accurate sampling without sign
problems. Thus, one can accurately determine the phase boundary
shifts at different $U$ beyond the mean-field level.  We find that
as $U$ increases, the critical value of $t_{3N}$ shifts towards a
larger value, thus effectively {\em stabilizing} the $Z_2$ TI phase.

\begin{figure}[t]
\epsfig{file=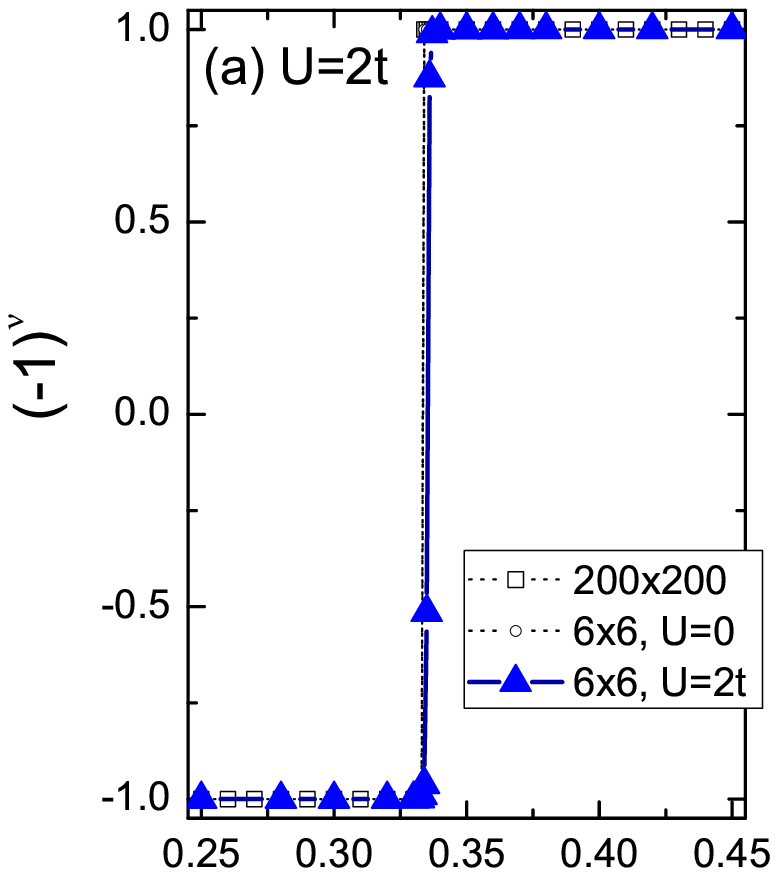,clip=0.1,width=0.45\linewidth,angle=0}
\epsfig{file=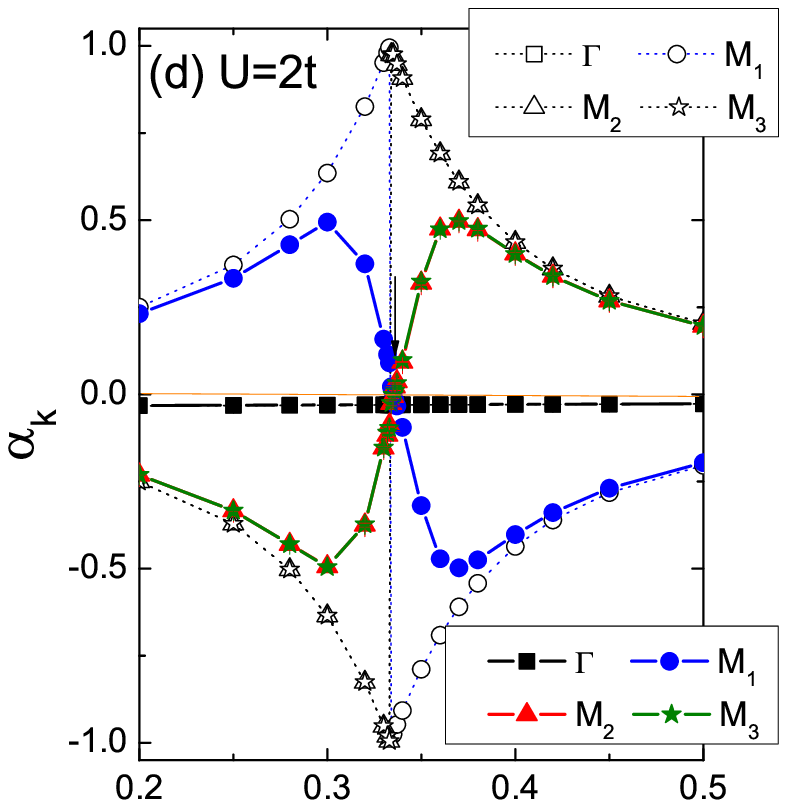,clip=0.1,width=0.51\linewidth,angle=0}
\epsfig{file=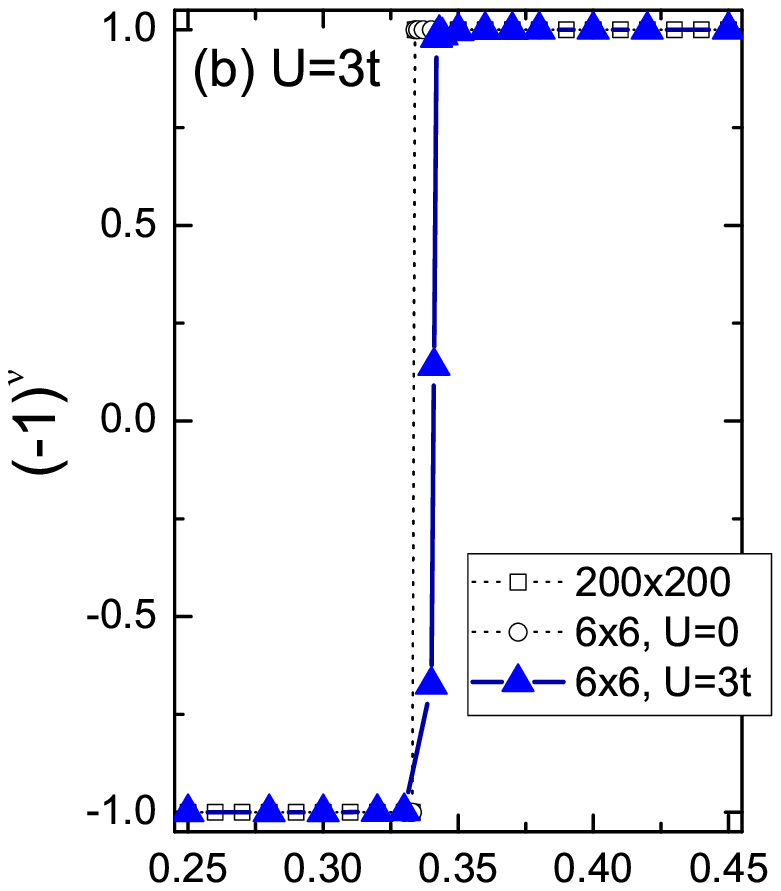,clip=0.1,width=0.45\linewidth,angle=0}
\epsfig{file=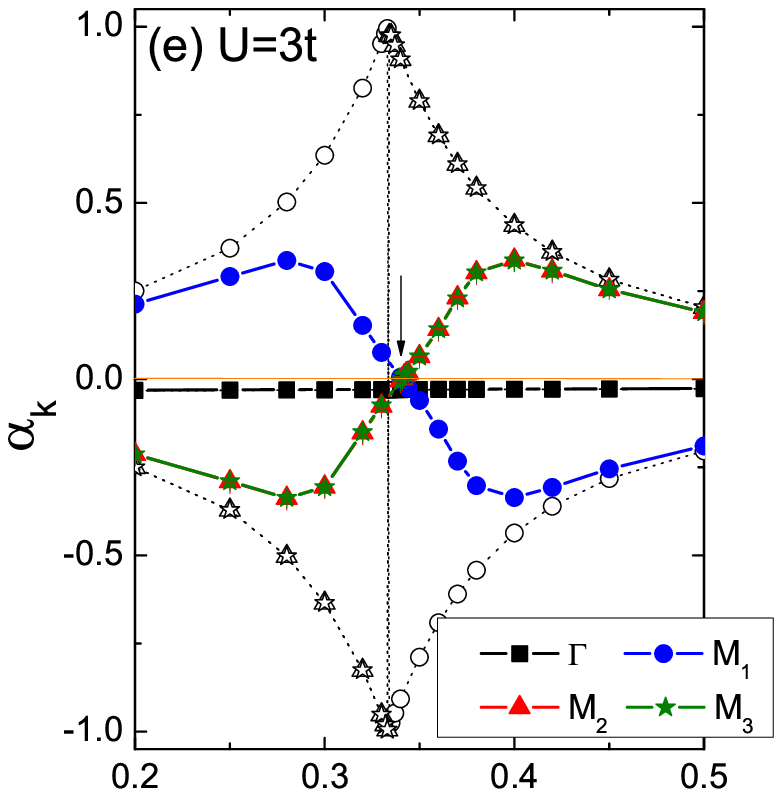,clip=0.1,width=0.52\linewidth,angle=0}
\epsfig{file=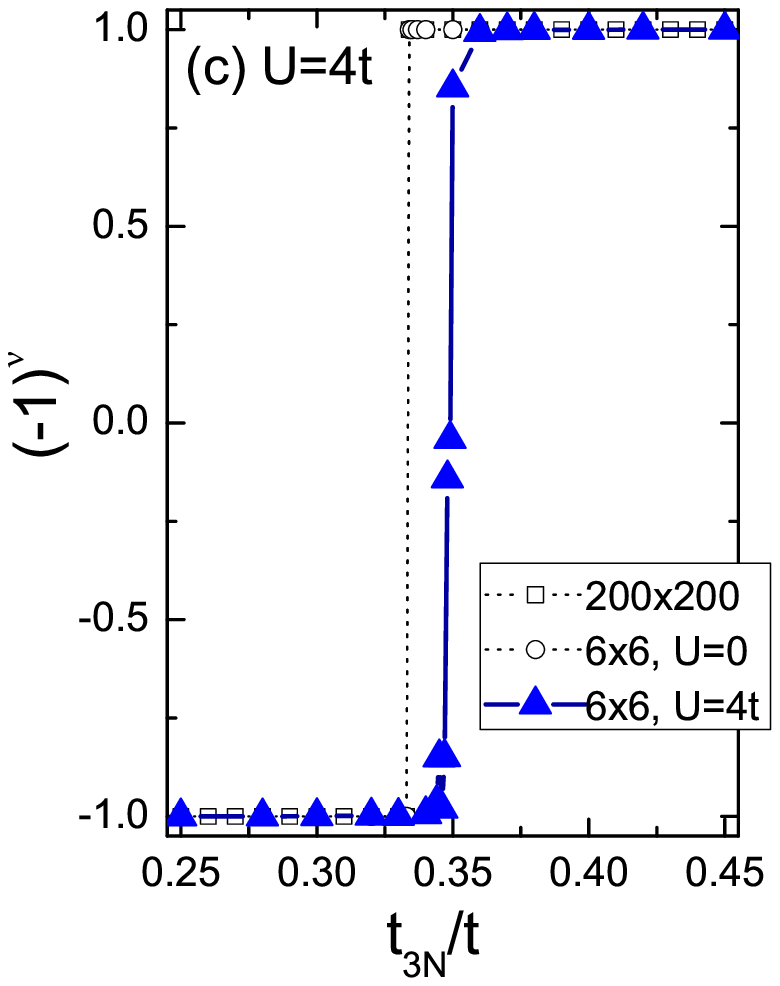,clip=0.1,width=0.45\linewidth,angle=0}
\epsfig{file=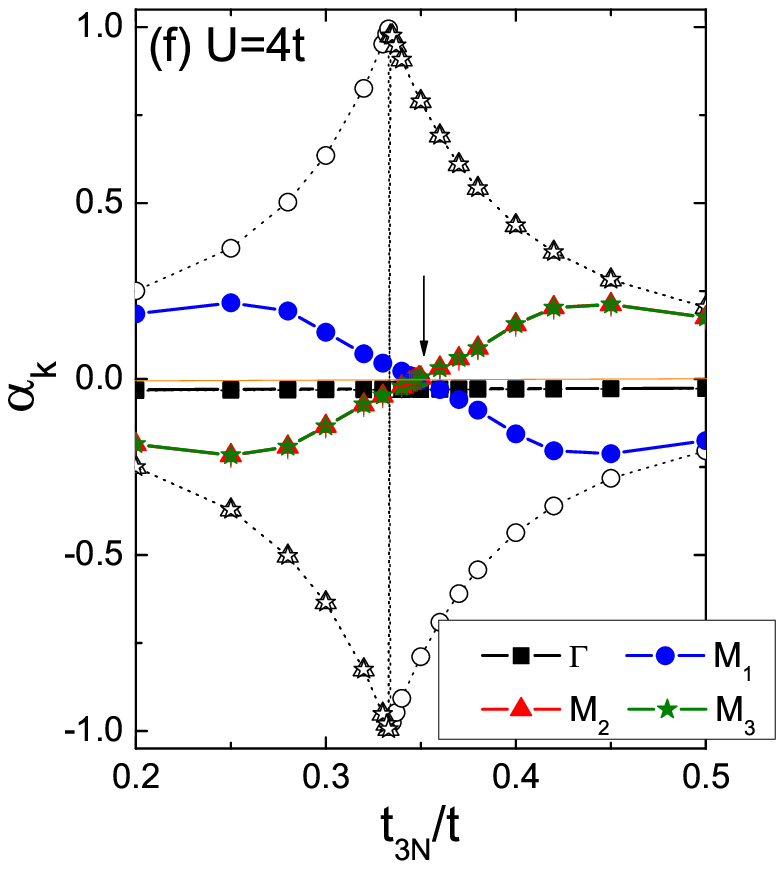,clip=0.1,width=0.52\linewidth,angle=0}
\caption{(Color online) (a)-(c) $Z_2$ invariant at $U/t=2$, $3$ and
$4$ vs $t_{3N}$. The spin-orbital coupling is $\lambda_{SO}=0.4t$.
The black squares show the $Z_2$ invariant given by the
tight-binding calculations with $200 \times 200$. The red circle
indicates the $Z_2$ invariant calculated by QMC simulations with $6
\times 6$ at $U=0$. The blue solid triangles depict the $Z_2$
invariant of the KMH model at $U \ne 0$. (d)-(f) show the
proportional coefficient $\alpha_{\mathbf k}$ determined by the
relation: $G_{\sigma}({\mathbf k}_{i},0)=\alpha_{\mathbf k_i}
\sigma^x$ from QMC simulations vs $t_{3N}$. All the open symbols
indicate noninteracting cases, i.e. $U=0$.  The solid symbols denote
interacting cases. } \label{fig:Z2_coeff}
\end{figure}

{\em Numerical Results.}-In our QMC calculations we use an imaginary
time step $\Delta\tau$ such that $\Delta \tau t=0.05$ and an inverse
temperature $\Theta$ such that $\Theta t=40$. For the noninteracting
case, for any finite $\lambda_{SO}$ and at $t_{3N} <t_c$, the system
is a $Z_2$ TI. We find that for $\lambda_{SO}=0.1t$, the model
transitions to a magnetic state at $U=3t$. To increase the threshold
value of $U$ needed to induce the magnetism, we consider a larger
$\lambda_{SO}=0.4t$ or even $\lambda_{SO}=t$ for different  $U$. For
comparison, in Fig. \ref{fig:Z2_coeff} we plot the value of the
$Z_2$ invariant as a function of $t_{3N}$ for different values of
$U$.  Open and solid symbols denote the noninteracting and
interacting cases, respectively. Unless otherwise stated, we
consider system sizes $L \times L=6 \times 6$ with periodic boundary
conditions. We also study the finite size effects on the topological
phase transition by comparing with $12 \times 12$ and $18 \times 18$
clusters.  We find negligible changes in the transition point for
these larger system sizes indicating that the location of the phase
transition is already accurately captured in the $L \times L=6
\times 6$ system size.

Using the single-particle Green's function we directly evaluate the
$Z_2$ invariant $\nu$,\cite{fu2007} where
\begin{eqnarray}
(-1)^{\nu}=\prod_{{\mathbf k_i} \in TRIM} \tilde{\eta}_{\mu_i},
\label{eqn:z2invariant}
\end{eqnarray}
and $\tilde{\eta}_{\mu_i}=\langle \tilde{\mu_i}
|P|\tilde{\mu_i}\rangle$ denotes the parity of the eigenstates of
zero-frequency Green's functions \cite{Wang:prx12} (see details in
supplementary information). Fig. \ref{fig:Z2_coeff} (a)-(c)
depict the dependence of the $Z_2$ invariant on $t_{3N}/t$ for
$U/t=2$, $3$ and $4$. The open black squares denote the $Z_2$
invariant given by tight-binding calculations with a $200 \times
200$ system size. The open red circles indicate the $Z_2$ invariant
calculated by QMC simulations for a $6 \times 6$ system at $U=0$.
The results are indistinguishable, confirming the accuracy of our
QMC calculations in the non-interacting limit, and validating the $6
\times 6$ system size results.  The location of topological phase
boundary is $t_{c}=\frac{1}{3}t$. In the TI phase, only the $M_1$
point is parity odd; the other three TRIM points are parity-even
(i.e. $\tilde{\eta}_{\Gamma}=\tilde{\eta}_{M_{2,3}}=+1$ and
$\tilde{\eta}_{M_{1}}=-1$), so $(-1)^{\nu}=-1$. Across the
transition upon increasing $t_{3N}$, $\tilde{\eta}_{M_{1,2,3}}$
change parity. $\Gamma$ and $M_1$ are parity-even whereas $M_{2,3}$
are parity-odd, so $(-1)^{\nu}=1$.

The blue solid triangles in Fig. \ref{fig:Z2_coeff} (a)-(c) depict
the dependence of the $Z_2$ invariant  on $t_{3N}$ for $U \ne 0$.
With correlations, the parity properties of the TRIM points still
remain and  Eq. (\ref{eqn:z2invariant}) to evaluate the $Z_2$
invariant is still valid.\cite{Wang:prx12,ara2012} Strictly
speaking, at each Monte Carlo measurement, the relation
$\tilde{\eta}_{\mathbf k}=\pm 1$ is not guaranteed. However, after a
thousand QMC samplings, $\langle \tilde{\eta}_{\mathbf k}
\rangle=\pm 1$ with tiny numerical errors. At weak interaction, the
phase boundary is barely seen to deviate. At $U=2t$, the phase
boundary is numerically estimated at $t_{3N}=0.335t$, which slightly
deviates from $t_{c}=\frac{1}{3}t$. By increasing $U$, however, one
can explicitly see that the interacting critical points not only
deviate from $t/3$ but move towards larger values, indicating the
topological phase is stabilized by interactions. At $U=3t$ and $4t$,
the topological phase transitions take place at $t_{3N}=0.341t$ and
$0.348t$, respectively. Moreover, when $\lambda_{SO}=t$, the
topological phase boundary at $U=6t$ occurs at $t_{3N}=0.352t$. This
indicates a significant ($\sim 10\%$) shift of the topological phase
boundary driven by the Hubbard interaction.  Moreover, no shift as a
function of $U$ is observed in a static Hartree-Fock mean-field
approximation. It is thus the quantum fluctuations originating in
the interactions that are important for shifting the phase boundary
and stabilizing the topological phase.  We believe this is likely to
be a rather general result.

Next, we investigate the single-particle Green's function in our
model. The parity operator is written as $\mathbb{I} \otimes
\sigma^x$, \cite{fu2007} and with inversion symmetry the Green's
functions for each spin are simply proportional to $\sigma^x$:
$G_{\sigma}({\mathbf k}_{i},0)=\alpha_{\mathbf k_i} \sigma^x$ [or
see Eq. (B2) in the supplemental information]. In Fig.
\ref{fig:Z2_coeff} (d)-(f) we show the proportionality coefficient
$\alpha_{\mathbf k}$ as a function of $t_{3N}$ for finite $U$. For
comparison, $\alpha_{\mathbf k}$ in the noninteracting case is also
depicted. At $U=0$, we find the universal relations,
\begin{eqnarray}
\alpha_{M_2}=\alpha_{M_3} \ \ \textrm{and} \ \
\alpha_{M_1}=-\alpha_{M_2},
 \label{eqn:kappa}
\end{eqnarray}
for all values of $\lambda_{SO}$ and $t_{3N}$. The values of
$\alpha_{\Gamma}$ behave smoothly as $t_{3N}$ is varied through
the topological critical points. However, the $\alpha$ coefficients
on the other TRIM points are divergent at $t_{3N}=t_{c}$ and
change sign at a topological phase transition. At
a critical point, the gap closes at the TRIM [c.f.
Fig. \ref{fig:bandstructure} (b)] so the zero-frequency Green's
functions are on the poles.\cite{gurarie2011} Irrespective of the
value of $\lambda_{SO}$, the location of the
 sign change is always at $t_{c}$,  consistent with the behavior of the $Z_2$ invariant.

Turning on the Hubbard interaction $U$, one can still observe the sign change in $\alpha_{\mathbf k}$ at the topological phase
transition. For finite $U$, the Green's functions retain their $\sigma^x$-like form and the universal relations in Eq. (\ref{eqn:kappa}) are still observed: $\alpha_{M_2} \simeq \alpha_{M_3}$ and $\alpha_{M_1} \simeq -\alpha_{M_2}$ within QMC simulation errors, independent of the value of $U/t$. However, the positions of $\alpha_{\mathbf k}$ begin to change their signs away from $t/3$, as indicated by arrows in Fig.\ref{fig:Z2_coeff} (d)-(f), which label the topological phase boundaries in the interacting case.
The locations for the sign change are consistent with the places where the $Z_2$ invariants dramatically jump. Note that at larger $U$ the magnitude of $\alpha_{\mathbf k}$ gradually vanish, but a sign change is still evident.

Also in Figs. \ref{fig:Z2_coeff} (d)-(f) one can observe how the $\alpha_{\mathbf k}$ coefficients evolve upon
increasing interactions.  In the noninteracting case, the coefficients flip sign dramatically at $t_{c}=t/3$.
However, the values of $\alpha_{\mathbf k}$ decrease by increasing $U$ and the sign-flip behavior becomes more smooth with stronger interaction. This corresponds to a smeared phase boundary indicated by the $Z_2$ invariant changes in Figs.\ref{fig:Z2_coeff} (a)-(c). Interestingly, away the topological phase transitions, e.g. $t_{3N}=0.2t$ and $0.5t$, the coefficients
$\alpha_{\mathbf k}$ for $U \ne 0$ seem to return to their noninteracting values. Therefore, interaction effects in
$\alpha_{\mathbf k}$ are most apparent as $t_{3N}$ approaches the topological phase transition points.

\begin{figure}[t]
\epsfig{file=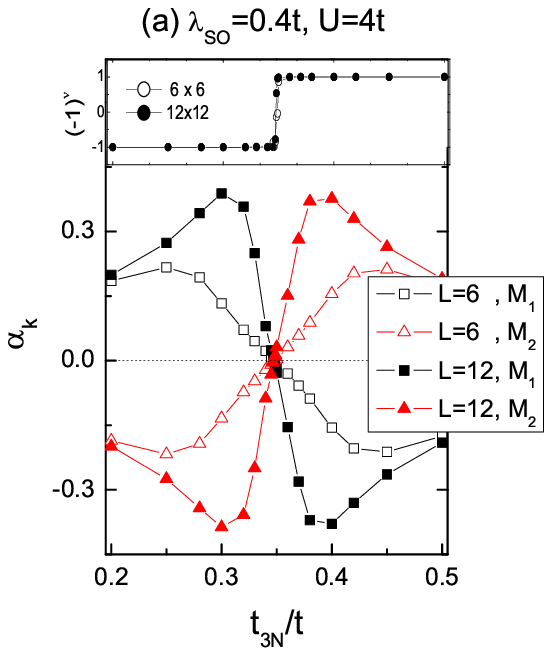,clip=0.1,width=0.555\linewidth,angle=0}
\epsfig{file=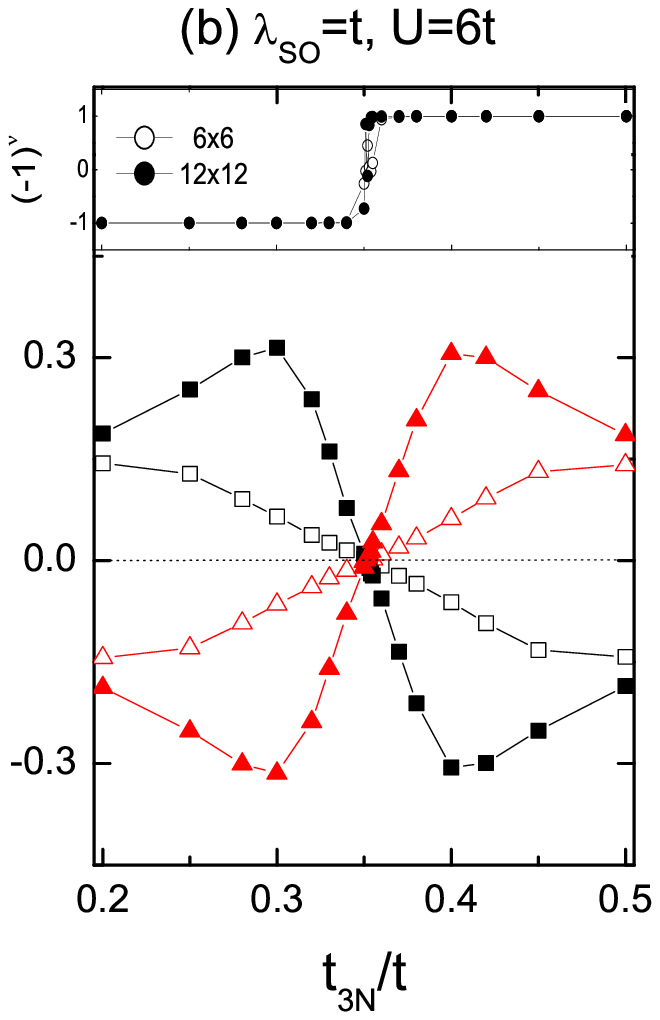,clip=0.1,width=0.425\linewidth,angle=0}
\caption{(color online) The comparison of the single-particle
Green's function coefficients $\alpha_{M_{1,2}}$ as a function of
$t_{3N}$ on $6\times 6$ (open symbols) and $12 \times 12$ (solid
symbols) for (a) $\lambda_{SO}=0.4t$ and $U=4t$ (b) $\lambda_{SO}=t$
and $U=6t$. The insets indicate the comparison of the $Z_2$
invariants vs $t_{3N}$ on the $6\times 6$ and $12 \times 12$
clusters with the same parameters.} \label{fig:Z2_finitesize}
\end{figure}

Finally, we investigate how finite size effects influence the
topological phase transition boundaries with finite $U$. For this
purpose, we compare the QMC results on $6 \times 6$ and $12 \times
12$ in Fig. \ref{fig:Z2_finitesize}. For a comparison for generic
parameters, we consider $\alpha_{\mathbf k}$ at the $M_1$ and $M_2$
points for (a) $\lambda_{SO}=0.4t$ and $U=4t$ and (b)
$\lambda_{SO}=t$ and $U=6t$. It is evident  that while stronger
interaction decreases $\alpha_{M_1}$ and $\alpha_{M_2}$ in
magnitude, the location of the sign change of $\alpha_{\mathbf k}$
barely depends on the system size. Independent of system size,
$\alpha_{M_1}$ and $\alpha_{M_2}$ switch sign at the same value ot
$t_{3N}$.  Such behavior shows that the topological phase transition
has a weak size dependence. The insets indicate the $Z_2$ invariant
for the two cases, also showing a small size dependence. However, on
a small size, a stronger $U$ [e.g. the inset of Fig.
\ref{fig:Z2_finitesize} (b)] will lead to a less sharp boundary
determined by the $Z_2$ invariants, compared to the $\alpha_{\mathbf
k}$ behavior. For the same numerical accuracy, one can investigate
the single-particle Green's functions on small sizes compared to the
$Z_2$ invariant to determine the topological phase transition
boundary.   This result implies that the single-particle Green's
function can be a powerful tool in detecting topological phase
transitions in interacting systems without the need to evaluate the
full topological invariant. (Although, this should certainly be
checked in a few cases as it is the precise quantity that is used to
distinguish the topological and non-topological phases.)

We note that the single-particle excitation gap is not a reliable
quantity to detect the topological phase boundary in finite-size
interacting systems. The single-particle gap should close when
undertaking the topological phase transition.
\begin{figure}[t]
\epsfig{file=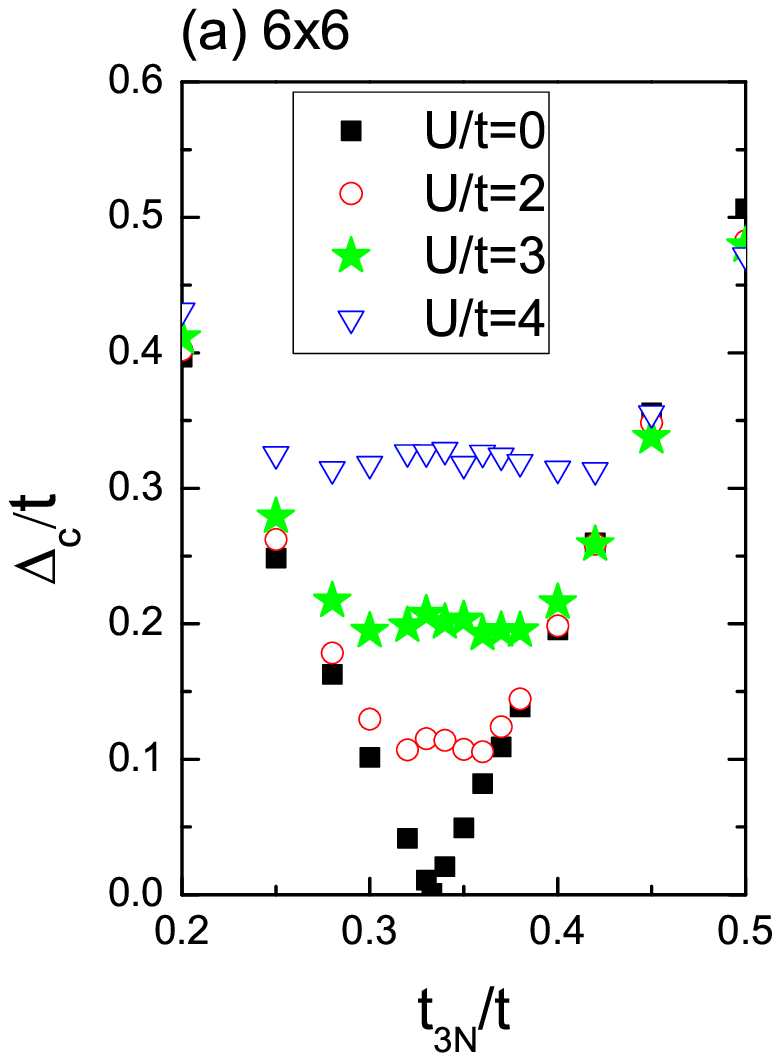,clip=0.1,width=0.5\linewidth,angle=0}
\epsfig{file=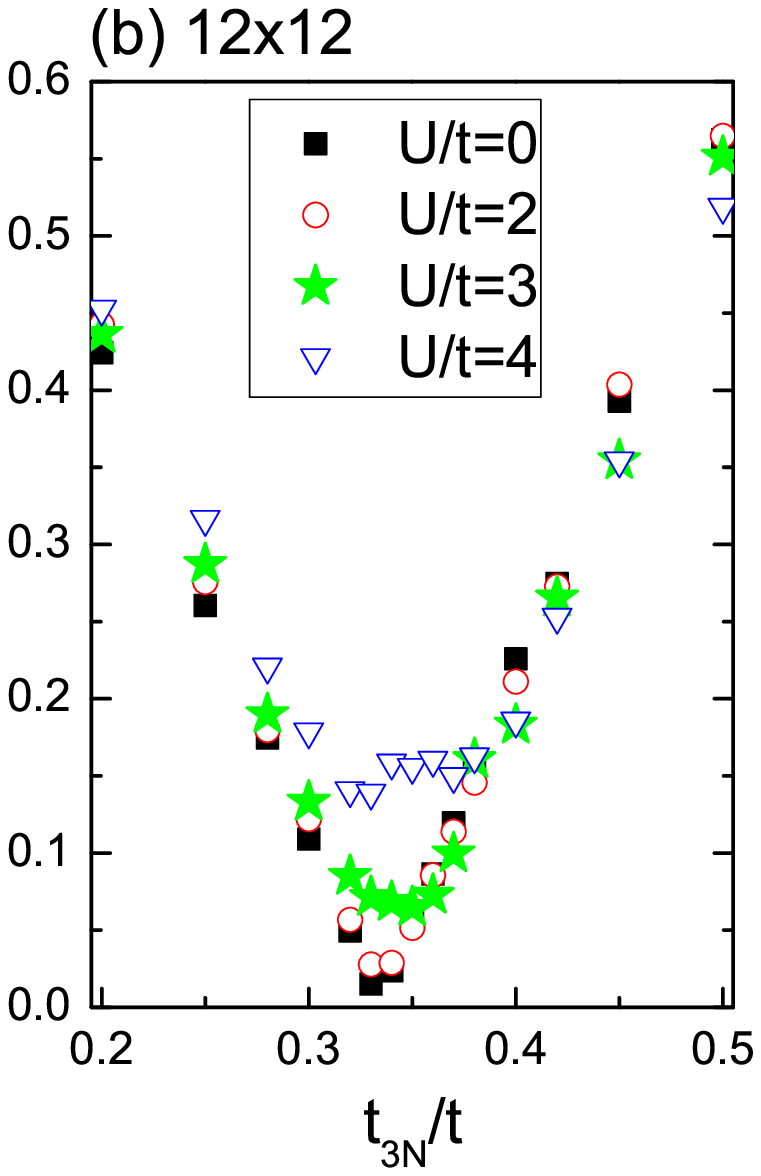,clip=0.1,width=0.435\linewidth,angle=0}
\caption{(Color online) Single-particle excitation gap $\Delta_c$
for different values of interaction: $U/t=0,2, 3$ and $4$ with
$\lambda_{SO}=0.4t$ on (a) $6\times 6$ and (b) $12\times 12$
clusters. For $U\neq0$ the single particle gap remains open across
the topological phase transition, in contrast to the behavior in a
non-interacting system.} \label{fig:chargegap}
\end{figure}
 As shown in Fig. \ref{fig:chargegap},
however, the single particle gaps are finite at the phase
transitions for $U\neq 0$ on the finite size simulations. Indeed,
comparing the $6\times 6$ and $12\times 12$ systems, we can clearly
see the decay tendency upon increasing size. The QMC results on
finite size scaling up to $18\times 18$ confirm that around the
phase boundaries the gaps vanish at $L\to \infty$. Thus the behavior
of the gaps is subject to strong finite size effect and the feature
of vanishing excitation can be only observed in the thermodynamic
limit. Moreover, the degree to which the phase transition is
obscured by the single particle gap also increases with increasing
$U$. With $12\times 12$, for $U=2t$ the gap seems to close around
$t_c$, but for $U=4t$ the behavior prevents one from determining the
topological phase boundary. Therefore, in an interacting system, one
should focus on the invariant itself and the
single-particle Green's function.

{\em Summary.}-We have studied a generalized Kane-Mele-Hubbard model
with an additional third-neighbor hopping term added.  In the
non-interacting limit the model exhibits a topological phase
transition as a function of third-neighbor hopping. By choosing
moderate Hubbard interactions without inducing antiferromagnetic
ordering, we study the topological phase transition in the
interacting level. Using a numerically exact, fermion sign-free
determinant projector QMC method, we have mapped the
interaction-dependence of this phase boundary. Our main result is
that interactions stabilize the topological phase by shifting the
phase boundary to enlarge the topological region. This effect is
absent in a static Hartree-Fock mean-field theory, which indicates
it is entirely the quantum fluctuations associated with the
interactions that enlarges the topological phase.  We also show that
the single-particle Green's function can more accurately determine
the phase boundary than the $Z_2$ invariant (which is derived from
it) for small system sizes. If this result can be reliably
generalized, this could be a useful insight in large-scale
``numerical searches" for real materials with topological
properties.  The importance of fluctuation effects in our model also
suggest that some density functional theory calculations could
incorrectly predict the topological invariant of materials where
quantum fluctuations are key to deciding the phase.

{\em Acknowledgements.}-HHH and GAF gratefully acknowledge financial
support through ARO Grant No. W911NF-09-1-0527, NSF Grant No.
DMR-0955778, and by grant W911NF-12-1-0573 from the Army Research
Office with funding from the DARPA OLE Program. LW thanks Matthias
Troyer for generous support. ZCG is supported in part by Frontiers
Center with support from the Gordon and Betty Moore Foundation.
Simulations were run on the Brutus cluster at ETH Zurich.



\begin{appendix}
\begin{widetext}

\section{Sign-free determinant projector QMC}

%
%

The determinant QMC has been shown to be an excellent and unbiased
approach to deal with strongly correlated system with Hubbard
interactions.\cite{sugiyama1986,sorella1989,white1989,scalettar1991,assaad1996,assaad1999,meng2010,cai2012}
In the projector algorithm, the ground state wave function
$|\Psi_0\rangle$ can be obtained using standard projection
procedures on a trivial wave function $|\Psi_T\rangle$, as long as
one requires $\langle \Psi_T |\Psi_0\rangle \ne 0$. The expectation
value of an observable $A$ is obtained by
\begin{eqnarray}
\langle A\rangle =\lim_{\Theta \to \infty}\frac{\langle \Psi_T |e^{-
\frac{\Theta }{2}H} A e^{- \frac{\Theta
}{2}H}|\Psi_T\rangle}{\langle \Psi_T |e^{-\Theta H}| \Psi_T\rangle}.
\label{eqn:expection}
\end{eqnarray}
The projection operator $e^{-\Theta H}$ can be discretized into many
time slices $e^{-\Theta H}=[e^{-\Delta \tau H}]^M$ with
$\Theta=\Delta \tau M$ where $\Delta \tau \ll 1$ and $M$ is the
number of time slices with a large integer number; $e^{-\Delta \tau
H}=e^{-\Delta \tau (H_0+H_U)}$ is the imaginary time-evolution
propagator during $\Delta \tau$. The noninteracting ground state of
$H_0$ is a good candidate for the trial wave function
$|\Psi_T\rangle$. With this trial wave function, we have confirmed
that the determinant projector QMC is in a good agreement with our
exact diagonalization results on a $L\times L=3\times 3$ system. By
the first order Suzuki-Trotter decomposition, one can decompose
$e^{-\Delta \tau H}$ as
\begin{eqnarray}
e^{-\Delta \tau H} \simeq e^{-\Delta \tau H_0}e^{-\Delta \tau H_U},
\label{eqn:suzuki}
\end{eqnarray}
where $H_0$ is the single-particle Hamiltonian of the generalized
Kane-Mele (KM) model as shown in Eq. (1) of the main text.
$H_U=\frac{U}{2}\sum_i(n_i-1)^2$ involves 4 fermionic operators and
cannot be represented in terms of single-particle basis. However, by
the discrete $SU(2)$-invariant Hubbard-Stratonovich transformation,
\cite{assaad1998} the interacting imaginary time-evolution operator
$e^{-\Delta \tau H_U}$ (for $U>0$) can be decomposed as
\begin{eqnarray}
e^{-\Delta \tau
\frac{U}{2}(n_i-1)^2}=\frac{1}{4}\sum_{l=\pm1,\pm2}\gamma(l)e^{i\sqrt{\Delta
\tau \frac{U}{2}}\eta(l)(n_i-1)}+O(\Delta
\tau^4),\label{eqn:HStransformation}
\end{eqnarray}
where $\gamma(\pm1)=1+\sqrt{6}/3$, $\gamma(\pm2)=1-\sqrt{6}/3$;
$\eta(\pm1)=\pm\sqrt{2(3-\sqrt{6})}$ and
$\eta(\pm2)=\pm\sqrt{2(3+\sqrt{6})}$ are 4-component auxiliary
fields determined by Monte Carlo samplings. Ref.
[\onlinecite{assaad2002,assaad2003,hung2011}] provide pedagogical
introductions about the QMC method. In this work, we employ $\Delta
\tau t=0.05$ in all the QMC simulations.

In the determinant algorithm with the Suzuki-Trotter decomposition
Eq. (\ref{eqn:suzuki}) and the Hubbard-Stratonovich transformation
Eq. (\ref{eqn:HStransformation}), the denominator of Eq.
(\ref{eqn:expection}) reads as
\cite{sorella1989,assaad2002,hohenadler2012,zheng2011} (up to a
constant factor)
\begin{eqnarray}
\langle \Psi_T | e^{-\Theta H} |\Psi_T\rangle &=& \langle \Psi_T |
\prod^M_{\tau=1} e^{-\Delta \tau H_{\tau}} |\Psi_T\rangle=\langle
\Psi_T | \prod^M_{\tau=1} e^{-\Delta \tau H_0}e^{-\Delta \tau
H_{U,\tau}} |\Psi_T\rangle\\ \nonumber &=&\sum_{\lbrace l_{i,\tau}
\rbrace} \Big\{ \prod_{i,\tau} \gamma(l_{i,\tau}) \prod_{\sigma} Tr
\Big(\prod^M_{\tau=1} e^{-\Delta \tau
\sum_{i,j}c^{\dag}_{i,\sigma}[{\bf
H^{\sigma}_{0}}]_{ij}c_{j,\sigma}}e^{i\sqrt{\Delta \tau
\frac{U}{2}}\eta(l_{i,\tau})(n_{i,\sigma}-\frac{1}{2})}
\Big)\Big\}\\
&=&\sum_{\lbrace l_{i,\tau} \rbrace} \Big\{ \prod_{i,\tau}
\gamma(l_{i,\tau}) p[\lbrace \eta(l_i) \rbrace]\Big\}, \nonumber
\end{eqnarray}
where $\sum_{l_{i,\tau}}$ runs over possible auxiliary
configurations $\eta(l_{i,\tau})$, where $i=1\sim N$, $\tau=1 \sim
M$; ${\bf H^{\sigma}_0}$ is the matrix kernel of $H_0$ with
spin-$\sigma$. The probability weight $p$ for an arbitrary auxiliary
configuration $\lbrace \eta(l_{i,\tau}) \rbrace$ is simply denoted
as \cite{hirsch1985}
\begin{eqnarray}
p(\lbrace \eta \rbrace)=\det \Big(O_{\uparrow}[\eta(l_{i,\tau})]
\Big)\det \Big(O_{\downarrow}[\eta(l_{i,\tau})] \Big),
\end{eqnarray}
where $\det \Big( O_{\sigma}[\eta(l_{i,\tau})] \Big)=Tr
\Big(\prod^M_{\tau=1} e^{-\Delta \tau
\sum_{i,j}c^{\dag}_{i,\sigma}[{\bf
H^{\sigma}_{0}}]_{ij}c_{j,\sigma}}e^{i\sqrt{\Delta \tau
\frac{U}{2}}\eta(l_{i,\tau})(n_{i,\sigma}-\frac{1}{2})} \Big)$. When
$p<0$, QMC simulations meet notorious minus-sign problems.

It has been proven that at half filling there exists a particle-hole
symmetry in the KM model such that the probability is always
positive-definitive.\cite{zheng2011,hohenadler2012} This character
still remains even considering the real-valued third-neighbor
hopping $t_{3N}$ in the generalized KM model. The particle-hole
transformation performs as
\begin{eqnarray}
c_{i,\sigma} \to \xi_i d^{\dag}_{i,\sigma}, \ \ c^{\dag}_{i,\sigma}
\to \xi_i d_{i,\sigma},\nonumber
\end{eqnarray}
where $\xi_i=-1$ ($\xi_i=1$) if $i$ belongs to $A$ ($B$) sublattice.
To show the positiveness of $p(\lbrace \eta \rbrace)$ in the
generalized KM model, we employ the particle-hole transformation on
the $t_{3N}$ hopping with $\downarrow$ but remain $\uparrow$
unchanged. Upon such a transformation, the $t_{3N}$ tight-binding
term turns out to be
\begin{eqnarray} & &-t_{3N}c^{\dag}_{i,\downarrow}c_{j,\downarrow} \nonumber \\
&\to& -t_{3N}\xi_i \xi_j
d_{i,\downarrow}d^{\dag}_{j,\downarrow}=-t_{3N}(-1)\xi_i \xi_j
d^{\dag}_{j,\downarrow}d_{i,\downarrow}.\nonumber
\end{eqnarray}
Note that the $t_{3N}$ hopping connects $A$ and $B$ lattices, so we
have $(-1)\xi_i \xi_j=1$. Therefore, upon the particle-hole
transformation,  $H^{\uparrow}_0$ and $H^{\downarrow}_0$ still have
identical matrix kernels.

The Hubbard interaction on $\downarrow$ transforms as
\begin{eqnarray}
& & i\sqrt{\Delta \tau
\frac{U}{2}}\eta(l_{i,\tau})(n_{i,\downarrow}-\frac{1}{2}) \nonumber \\
&\to& i\sqrt{\Delta \tau
\frac{U}{2}}\eta(l_{i,\tau}) \Big\{ (\xi_i)^2 d_{i,\downarrow}d^{\dag}_{i,\downarrow}-\frac{1}{2} \Big\} \nonumber \\
&=&-i\sqrt{\Delta \tau
\frac{U}{2}}\eta(l_{i,\tau})(d^{\dag}_{i,\downarrow}d_{i,\downarrow}-\frac{1}{2}),\nonumber
\end{eqnarray}
which is the complex conjugate of $H_U$ on $\uparrow$. Consequently,
upon the particle-hole symmetry, one can have $\det(O_{\downarrow})=
\det( O_{\uparrow} )^*$ and the probability weight
$p=\det({O_{\uparrow}})\det({O_{\downarrow}})=|\det(O_{\uparrow})|^2$
being real positive. The QMC simulation in the half-filled
generalized KM model is sign-free and numerically exact.

\section{single particle Green's functions}

Without sign problems, the QMC samplings provide highly accurate not
only equal-time Green's functions but also time-displaced Green's
functions in real space \cite{assaad1996,feldbacher2001}
\begin{eqnarray}
G_{\sigma}(\vec{r},\tau)=\langle \Psi_0 |
c_{\sigma}(\vec{r},\tau)c^{\dag}_{\sigma}(0)|\Psi_0\rangle,\nonumber
\end{eqnarray}where $\tau >0$. By performing double
 Fourier transformation we obtain the Green's functions
in momentum space and Matsubara frequency, i. e.
$G_{\sigma}(\mathbf{k},i\omega_{n})$.

\begin{figure}[htb!]
\epsfig{file=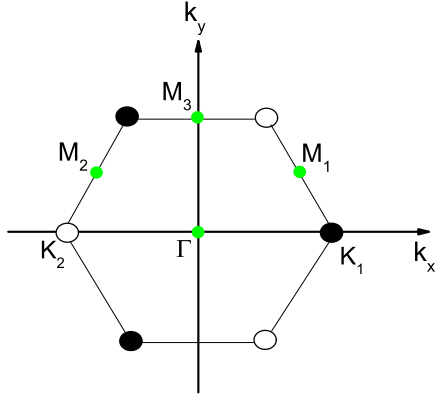,clip=0.1,width=0.32\linewidth,angle=0}
\caption{(Color online)  The Brillouin zone of the honeycomb
lattice. The time-reversal invariant momentum (TRIM) points labeled
by the green dots are $\Gamma=(0,0)$,
$M_{1,2}=(\pm\frac{\pi}{\sqrt{3}a},\frac{\pi}{3a})$, and
$M_{3}=(0,\frac{2\pi}{3a})$. The open and solid circles denote
graphene Dirac points $K_{1,2}=(\pm\frac{4\pi}{3\sqrt{3}a},0)$}.
\label{fig:bandstructure2}
\end{figure}
It has been shown that zero frequency Green's functions are able to
evaluate the  $Z_2$ invariant index in the interacting
case.\cite{wang2012} The $Z_2$ invariant is determined by the parity
of the eigenvectors of the inverse Green's functions
\begin{eqnarray}
[G({\mathbf k}_{i},0)]^{-1}|\mu_i \rangle=\mu_i
|\mu_i\rangle.\nonumber
\end{eqnarray}
Note that since there still exists an inversion symmetry in the
generalized KMH model, the inverse Green's functions and the parity
operator have simultaneous eigenvectors, \textit{i.e.}
$P|\mu_{i}\rangle=\eta_{\mu_{i}}|\mu_{i}\rangle$. In the
(generalized) KM model,  the parity operator exchanges $A$, $B$
sublattices independent of spin index. Therefore, with the spinor
convention $\Psi^{\dag}=(c^{\dag}_{A,\uparrow} \
c^{\dag}_{B,\uparrow} \ c^{\dag}_{A,\downarrow} \
c^{\dag}_{B,\downarrow})$, the parity operator is defined as $P=I
\otimes \sigma^{x}$.\cite{fu2007} In the QMC simulations, the
particle-hole symmetry provides $G_{\uparrow}({\mathbf
k_i},0)=G_{\downarrow}({\mathbf k_i},0)$ while  ${\mathbf k}_{i}$ is
time-reversal invariant momentum (TRIM), i.e. ${\mathbf k}=-{\mathbf
k}$. Therefore, we can directly diagonalize $G_{\sigma}({\mathbf
k}_{i},0) =[-H_{\mathbf k}-\Sigma({\mathbf k}_{i},0)]^{-1}$ instead
of inverse Green's functions for all ${\mathbf k}_{i}\in$ TRIM
points
\begin{eqnarray}
G_{\sigma}({\mathbf k}_{i},0)|\tilde{\mu_i} \rangle=\tilde{\mu_i}
|\tilde{\mu_i}\rangle \nonumber,
\end{eqnarray} and choose the eigenvectors associated with positive
eigenvalues ($\tilde{\mu_i}>0$, denoting occupied bands and are
called right-zero \cite{Wang:prx12}). In the honeycomb lattice, the
TRIM points are $\Gamma$, $M_{1,2,3}$ as depicted in Fig.
\ref{fig:bandstructure2}. Then we can employ the formalism proposed
by Fu and Kane\cite{fu2007,Wang:prx12} to identify the $Z_2$
invariant as
\begin{eqnarray}
(-1)^{\nu}=\prod_{{\mathbf k_i} \in TRIM} \tilde{\eta}_{\mu_i},
\end{eqnarray}
where $\tilde{\eta}_{\mu_i}=\langle \tilde{\mu_i}
|P|\tilde{\mu_i}\rangle$. When $\nu=0$ for trivial insulator,
whereas $\nu=1$ is a $Z_2$ topological insulator. In the case of
$U=0$, $\tilde{\eta}_{\mu_i}=\pm1$. In the cases of finite $U$, we
find that $\langle \tilde{\eta}_{\mu_i} \rangle=\pm1$ can be still
obtained by sufficient QMC simulations. As $t_{3N}$ approaches the
topological critical point, $(-1)^{\nu}$ will be smeared out and is
laid between $\pm1$. In this case, more QMC samplings are required
for more accurate values.

Note that since  $G_{\uparrow}({\mathbf
k}_{i},0)=G_{\downarrow}({\mathbf k}_{i},0)$, and $G({\mathbf
k}_{i},0)$ [$=G_{\uparrow}({\mathbf k}_{i},0) \oplus
G_{\downarrow}({\mathbf k}_{i},0)$] and $P$ ($=I \otimes \sigma^x$)
have the simultaneous eigenvector sets, one has a relation:
\begin{eqnarray}
G_{\sigma}({\mathbf k}_{i},0)=\alpha_{\mathbf k_i} \sigma^x.
\label{eqn:gamma}
\end{eqnarray}
In the context we show that in addition to the $Z_2$ invariant, the
proportional coefficient $\alpha_{\mathbf k}$ also plays another
role to characterize a topological phase transition and even is more
sensitive than $\nu$ numerically.  Upon the topological phase
transition, the bulk gap will close at the TRIM points. Thus the
single-particle Green's functions will be divergent on the poles.
\cite{gurarie2011}

The relation Eq. (\ref{eqn:gamma}) should be expected both in the
noninteracting and interacting cases. However, as $U \ne 0$ Eq.
(\ref{eqn:gamma}) is not guaranteed in a single measurement in the
QMC simulations. The proportionality relation between the
zero-frequency Green's functions and the parity matrix $\sigma^x$
can be recovered only upon enough samplings. To interpret this, we
present the $6 \times 6$ benchmark results for the matrix elements
of the zero-frequency Green's functions at ${\mathbf k=M_1}$ as a
function of the number of measurements in Figs. \ref{fig:benchmark}.
\begin{figure}[!htb]
\epsfig{file=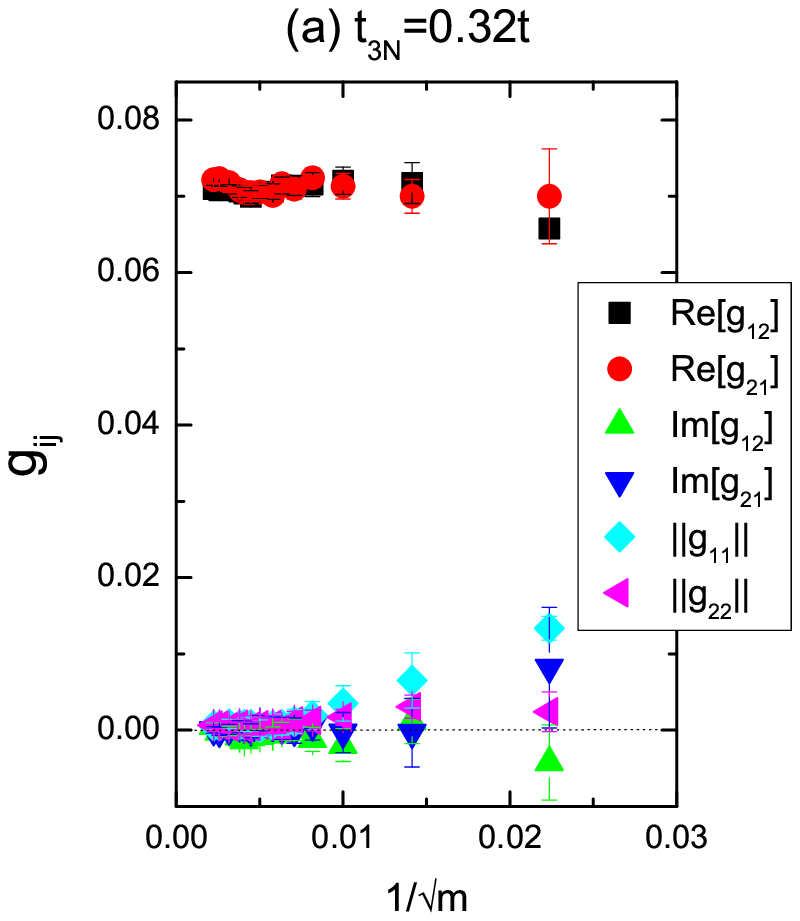,clip=0.1,width=0.3\linewidth,angle=0}
\epsfig{file=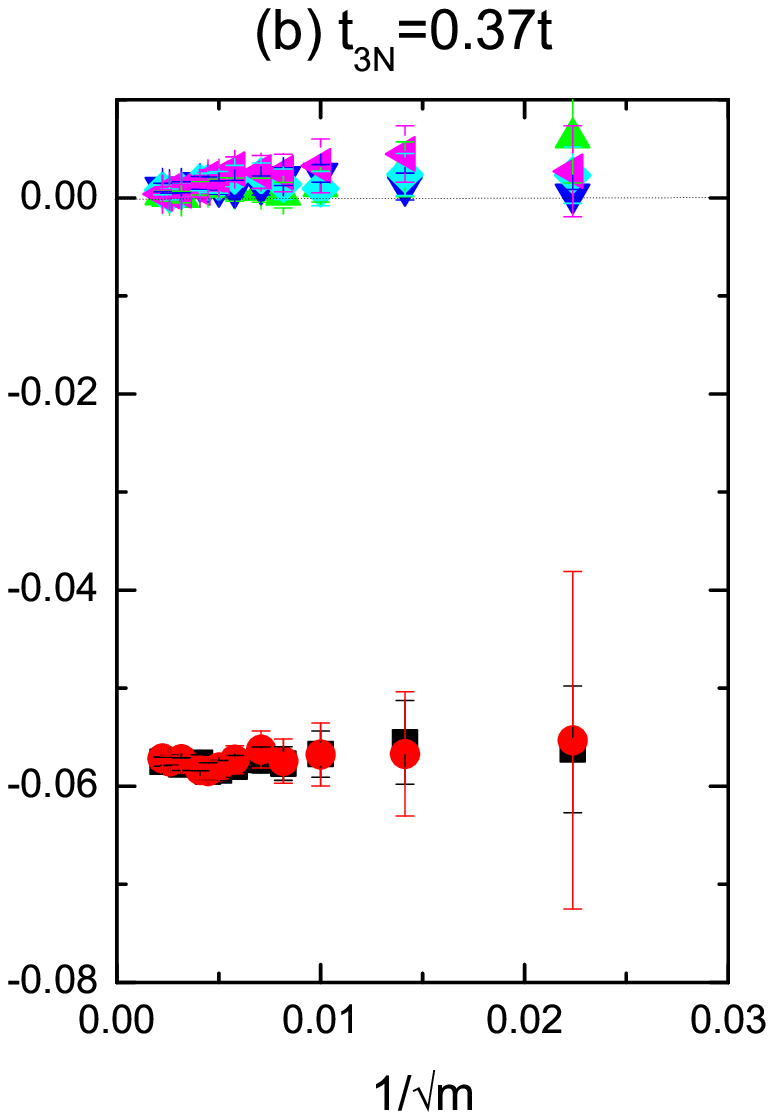,clip=0.1,width=0.24\linewidth,angle=0}
\caption{(Color online) The matrix elements of the zero-frequency
Green functions $G({\mathbf M_1},0)$ vs the number of samplings $m$
at (a) $t_{3N}=0.32t$ and (b) $t_{3N}=0.37t$. $\lambda_{SO}=0.4t$
and $U=4t$. $\textrm{Re}[g_{ij}]$ and $\textrm{Im}[g_{ij}]$ denote
the real part and imaginary part of $[G({\mathbf M_1},0)]_{ij}$,
respectively; $||g_{ii}||$ denotes the diagonal component of
$G({\mathbf M_1},0)$ in magnitudes.} \label{fig:benchmark}
\end{figure}
$g_{ij}=[G({\mathbf M_1},0)]_{ij}$ and $m$ denotes the number of
measurements. $\lambda_{SO}=0.4t$ and $U=4t$ are used. In this case
the topological phase boundary is identified at $t_{3N}=0.348t$. We
choose the value of $t_{3N}$ close to the critical point. Fig.
\ref{fig:benchmark} (a) shows $t_{3N}=0.32t$ in the $Z_2$
topological insulator phase and (b) for $t_{3N}=0.37t$ in the
trivial insulator. From the panels, it is evident that the structure
of the Green's function does not fit Eq. (\ref{eqn:gamma}) as there
are no sufficient samplings. At small $m$, the real parts of
$g_{12}$ and $g_{21}$ are not equal; furthermore $g_{12}$ and
$g_{21}$ have imaginary parts and $g_{11}$ and $g_{22}$ are finite.
However, one can see that upon sampling sufficient times
$\textrm{Re}[g_{12}] \simeq \textrm{Re}[g_{21}]$, and meanwhile
$\textrm{Im}[g_{12}]$, $\textrm{Im}[g_{21}]$, $||g_{11(22)}||$ go to
zero. Thus in the $m \to \infty$ limit, Eq. (\ref{eqn:gamma}) is
recovered. Also note that $\alpha_{\mathbf M_1}=\textrm{Re}[g_{12}]$
in both cases indicate opposite sign as observed by the signature of
the topological phase transition. Moreover, by such $m$ scaling, we
also confirm that the value of the $Z_2$ invariant also shows
monotonically close to $\pm 1$. In our paper we choose the value of
$m$ large enough to determine the $\sigma^x$ structure and extract
the coefficients.

\section{Critical Hubbard interactions for antiferromagnetism}

In the generalized KMH model  a strong Hubbard interaction can also
derive the antiferromagnetic (AF) ordering, due to the bipartite
lattice structure. Similarly to the KMH model (with $t_{3N}=0$)
\cite{rachel2010,zheng2011,hohenadler2011},  in the generalized KMH
model, finite values of $\lambda_{SO}$ also break the $SU(2)$
symmetry down to the  $U(1)$ symmetry and the dominant magnetism
behavior lies on x-y plane. The planar spin structure factor can be
defined as\cite{hohenadler2011,zheng2011}
\begin{eqnarray}
S_{AF}=\sum_{\vec{r},\vec{r}_j}(-1)^{\vec{r}_i+\vec{r}_j}\langle
S^{+}_iS^-_j +S^{-}_iS^+_j\rangle.\nonumber
\end{eqnarray}
$(-1)^{\vec{r}_i}=1(-1)$ for $i \in A(B)$ sublattice. This is
similar to determining the N\'eel type ordering in a square lattice
by using the antiferromagnetic spin structure factor at ${\mathbf
k}=(\pi,\pi)$.
\begin{figure}[!tb]
\epsfig{file=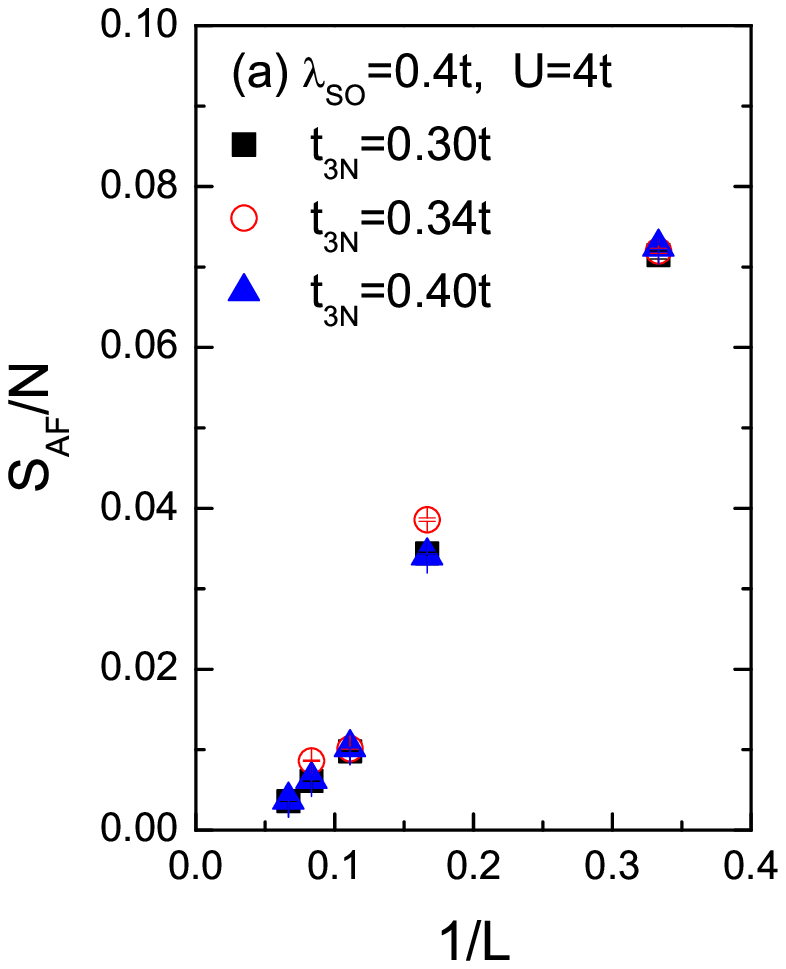,clip=0.1,width=0.235\linewidth,angle=0}
\epsfig{file=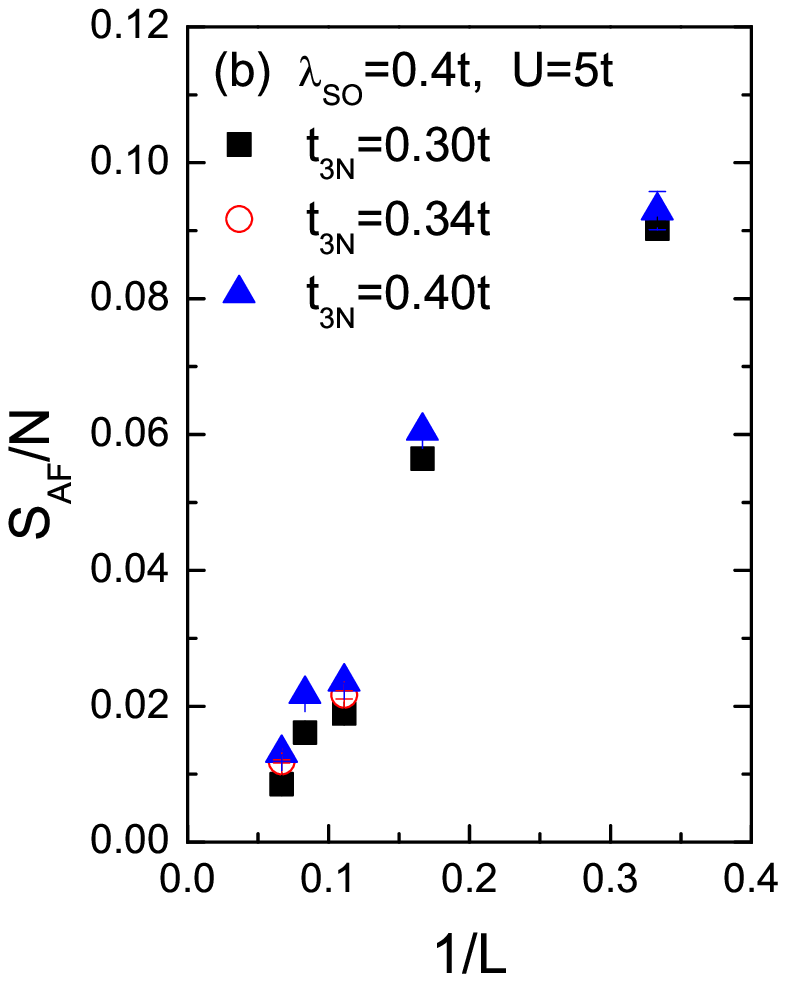,clip=0.1,width=0.23\linewidth,angle=0}
\epsfig{file=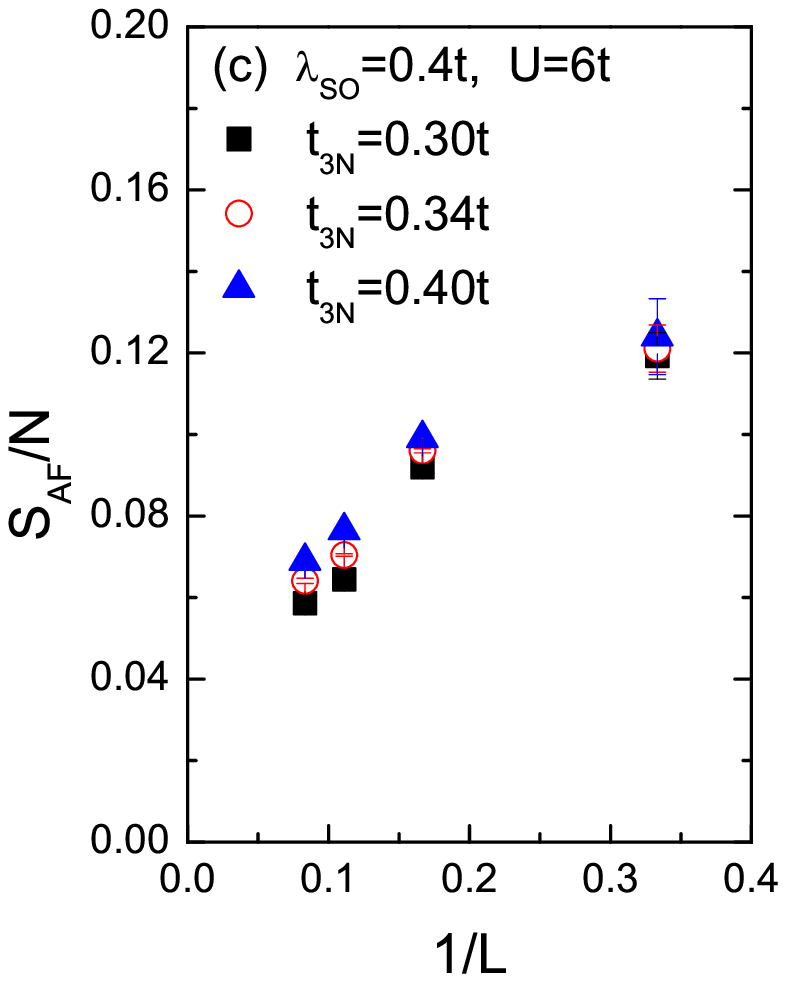,clip=0.1,width=0.23\linewidth,angle=0}
\epsfig{file=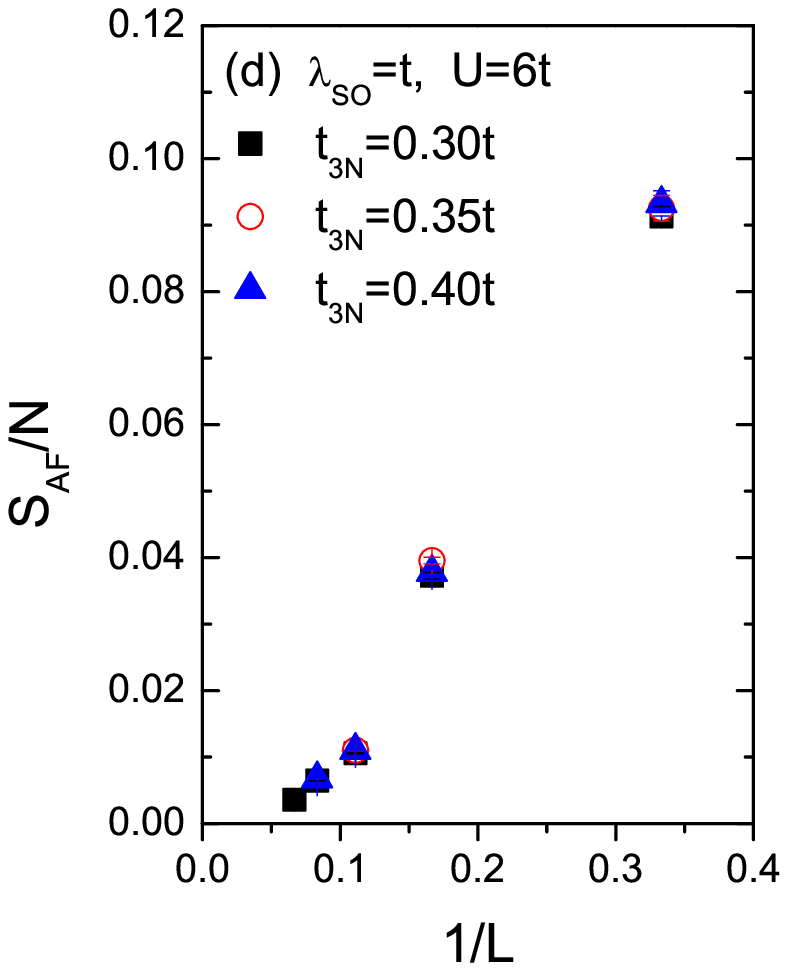,clip=0.1,width=0.235\linewidth,angle=0}
\caption{(Color online) (a)-(c) The finite size scaling of the
antiferromagnetic spin structure factor $S_{AF}/N$ vs $1/L$ at
$\lambda_{SO}=0.4t$ and different $U=4t,5t,6t$. (d) $S_{AF}/N$ vs
$1/L$ at $\lambda_{SO}=t$ and $U=6t$. Here $N=2\times L^2$.}
\label{fig:criticalU}
\end{figure}
To identify whether there exists the antiferromagnetism in the
thermodynamic limit, we study the finite size scaling behavior of
$S_{AF}$ at $L \to \infty$. Generally speaking, the spin-orbital
coupling will suppress AF ordering and larger $\lambda_{SO}$'s are
associated with larger $U_c$'s to induce the AF ordering. Note that
due to the presence of third nearest neighboring hopping $t_{3N}$
which favors the N\'eel pattern in the second order perturbation,
the threshold interaction $U_c$ in the generalized KMH model is
smaller than that in the KMH model.

The QMC results on $S_{AF}/N$ vs $1/L$ are shown in Figs.
\ref{fig:criticalU}. In (a), we can see that, for
$\lambda_{SO}=0.4t$,
 $U=4t$ is not sufficiently large to induce the AF
ordering. At $U=5t$, $S_{AF}$ is enhanced and the $U$ value is close
to the critical value to drive the AF ordering. In (c) under the
interaction $U=6t$, $S_{AF}$ saturates to finite values at $1/L \to
0$, suggesting that the AF ordering exists in the thermodynamic
limit. Fig. \ref{fig:criticalU} (d) depicts the case of $U=6t$ but
at $\lambda_{SO}=t$. Compared to (c), where an AF ordering is
induced, the structure factor in (d) still goes to zero in the $L
\to \infty$ limit. Thus, stronger spin-orbital couplings obviously
suppress the existence of AF ordering and raise values of critical
interactions in the generalized KMH model.

\section{Single-particle excitation}

In this subsection, we present the approach to evaluate  the
single-particle excitation (charge gap) $\Delta_c$ in the QMC
simulations. The charge gap is defined as the energy cost to add a
particle into (or remove a particle from) the system composed of $N$
particles. Assuming that we have $\hat{H} |
\Psi^{N+1}_n\rangle=E^{N+1}_n | \Psi^{N+1}_n\rangle$ and $\hat{H} |
\Psi^{N}_n\rangle=E^{N}_n | \Psi^{N}_n\rangle$, then the charge gap
reads $\Delta_c \equiv E^{N+1}_0-E^{N}_0$. It can be obtained via
calculating the on-site time-displaced Green's functions which are
written as
\begin{eqnarray}
G(\vec{r}=0,\tau)&=&\frac{1}{N}\sum_{i,\sigma}G_{\sigma}(i,i;\tau)
\nonumber \\ &=&\frac{1}{N}\sum_{i,\sigma}\langle \Psi^N_0 |
c_{\sigma}(i,\tau)c^{\dag}_{\sigma}(i)|\Psi^N_0 \rangle \nonumber \\
&=& \frac{1}{N}\sum_{i,\sigma}\langle \Psi^N_0 | e^{\tau
\hat{H}}c_{\sigma}(i)e^{-\tau \hat{H}}c^{\dag}_{\sigma}(i)|\Psi^N_0 \rangle. \nonumber \\
&=& \frac{1}{N}\sum_{n,i,\sigma} e^{-\tau (E^{N+1}_n-E^N_0)}
|\langle \Psi^N_0 | c_{\sigma}(i) |\Psi^{N+1}_n \rangle|^2.\nonumber
\end{eqnarray}
Therefore, at large $\tau$, we have $G(\vec{r}=0,\tau) \sim e^{-\tau
\Delta_c}$
 and
then one can find the slope of  $\ln{G(\vec{r}=0,\tau)}$ at large
$\tau$ to determine the value of $\Delta_c$. Refs.
[\onlinecite{assaad1996,feldbacher2001,assaad1996b}] and
[\onlinecite{hung2011}] provide the detailed descriptions. The
evaluation of the excitation by the on-site single-particle Green's
function can determine the value of the single-particle excitation
without concerning about specific momentum points, e.g.
$\Delta_c({\mathbf k})$. (Note that the gap of the KM model with
$\lambda=0$ closes at the Dirac points $K_{1,2}$, whereas the gap of
the generalized KM model with $t_c$ closes at $M_{1,2,3}$.)

\end{widetext}
\end{appendix}

\end{document}